\documentclass{article}

\usepackage{arxiv}

\usepackage[utf8]{inputenc} 
\usepackage[T1]{fontenc}    
\usepackage{hyperref}       
\usepackage{url}            
\usepackage{booktabs}       
\usepackage{amsfonts}       
\usepackage{nicefrac}       
\usepackage{microtype}      
\usepackage{lipsum}
\usepackage{graphicx}
\usepackage{amsmath}
\usepackage{amssymb}
\usepackage{authblk}

\title{Complete multipolar description of reflection and transmission across a metasurface for perfect absorption of light}

\author{Romain Dezert, Philippe Richetti, and Alexandre Baron}
 \affil{Univ. Bordeaux, CNRS, Centre de Recherche Paul Pascal, UMR5031, 33600 Pessac, France}
 \affil{email: alexandre.baron@u-bordeaux.fr}

\begin{document}
\maketitle

\begin{abstract}
Relating the electromagnetic scattering and absorption properties of an individual particle to the reflection and transmission coefficients of a two-dimensional material composed of these particles is a crucial concept that has driven both fundamental and applied physics. It is at the heart of both the characterization of material properties as well as the phase and amplitude engineering of a wave. Here we propose a multipolar description of the reflection and transmission coefficients across a monolayer of particles using a vector spherical harmonic decomposition. This enables us to provide a generalized condition for perfect absorption which occurs when both the so-called \textit{generalized Kerker condition} is reached and when the superposition of odd and even multipoles reaches a critical value. Using these conditions, we are able to propose two very different designs of two-dimensional materials that perfectly absorb a plane electromagnetic wave under normal incidence. One is an infinite array of silica microspheres that operates at mid-infrared frequencies, while the other is an infinite array of germanium nano-clusters that operates at visible frequencies. Both systems operate in a deeply multipolar regime. Our findings are important to the metamaterials and metasurfaces communities who design materials mainly restricted to the dipolar behavior of individual resonators, as well as the self-assembly and nanochemistry communities who separate the individual particle synthesis from the materials assembly.
\end{abstract}


\section{Introduction}

When an electromagnetic wave impinges on a single layer of particles or atoms, local currents are generated within the particles that in turn radiate a field that will interfere with the incoming wave to produce a reflection and a transmission \cite{jackson1999classical}. The transmission and reflection coefficients, also known as the \textit{scattering parameters}, of a two-dimensional material are essential observables in physics and chemistry. They are often used to characterize the effective  properties of a material. They may also be engineered to target a desired application. In recent years, many two-dimensional materials composed of sub-wavelength polarizable units or scatterers have been designed and modelled to generate extraordinary and tailored reflection and transmission properties. The recent surge in metasurfaces is a good example of arrays of particles designed to control waves both in amplitude and in phase for wave-front shaping \cite{arbabi2015dielectric,decker2015high,khorasaninejad2017metalenses}, filtering \cite{glybovski2016metasurfaces} or perfect absorption \cite{Alaee_2017,thongrattanasiri2012complete,liu2017experimental}.

The field radiated by subwavelength individual systems is often decomposed into multipoles. Very recently, multipolar decompositions have become an essential tool in nanophotonics and metamaterials engineering. Alaee et \textit{al}. provide an interesting perspective on the matter in \cite{alaee2019exact}. Here we shall consider the multipolar decomposition on a vector spherical harmonic basis. This is convenient as it reduces the problem to the sole determination of the multipolar scattering coefficients $a_{n,m}$ and $b_{n,m}$ that fully describe how the field is emitted \cite{jackson1999classical,bohren2008absorption,grahn2012electromagnetic}. Numerous problems requiring the expression of the field radiated by an infinite array of particles have been treated in the past. 
As early as 1932, Strachan tried to develop a theory of the light reflected by a monomolecular layer \cite{strachan1933reflexion}, though we now know that his theory was incomplete as it neglected the magnetic polarization density \cite{kuester2003averaged}. Later arrays were mostly described in terms of electric and magnetic effective dipole moments \cite{tretyakov2003analytical, evlyukhin2010optical,bowen2017effective}, but also recently including multipolar contributions up to the octupoles \cite{savinov2014toroidal,terekhov2019multipole}. These works are based on expressing the moments in cartesian coordinates but are usually quite involved as the expansions become complex for increasing multipole orders. For this reason, they are usually truncated at low orders, which potentially limits the domain of application of these methods. We may wonder why cartesian coordinates are used in metasurface analysis, since multipolar decompositions up to any order using vector spherical harmonics are widely used and adapted to the study of individual particles. 

In this work and in contrast to what has been done thus far, we directly relate the complex reflection and transmission coefficients of a two-dimensional array composed of single subwavelength scatterers of arbitrary shape to the multipolar coefficients of the particles in the array. These coefficients can be directly computed from the currents generated inside a single particle of the array using Grahn's decomposition \cite{grahn2012electromagnetic}. These currents can be calculated by various numerical methods. This formalism is simple, powerful, totally general and gives a lot of physical insight on the interaction of light with particles of arbitrary complexity. To demonstrate its use, we apply the formalism to absorption, which enables us to derive two generalized conditions that provide the basic recipe to achieve total absorption of an electromagnetic plane wave under normal incidence by an array of multipolar particles. It should be noted that both these conditions have already been given in the literature for arrays composed of dipoles \cite{ra2015full,Alaee_2017}. However, here it is generalized. It turns out that many multipolar systems may act as perfect absorbers which is a definite improvement made possible by our formalism. We are able to provide a generalized interpretation of critical coupling in terms of spherical multipoles as well as provide two examples of perfect absorbing infinite arrays based on very different systems: single silica microparticles that absorb in the far-infrared due to phonon excitations and clusters of germanium particles that absorb at optical frequencies. In this last case, the absorption is obtained via an engineering of the multipolar behavior tailored with the cluster geometry. 

\section{Theoretical formalism}

\begin{figure}[t!]
   \centering
   \includegraphics[width=0.7\textwidth]{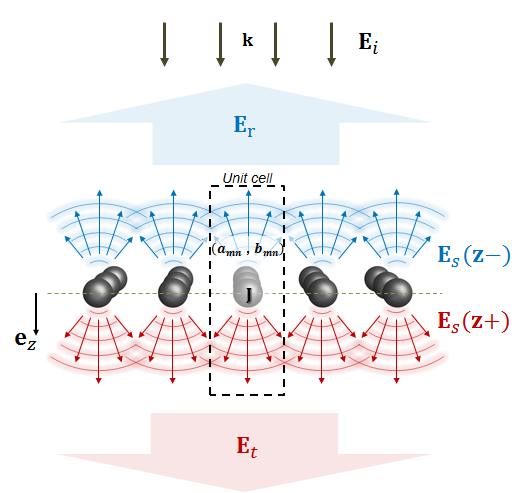}
   \caption{Artistic rendition of the multipolar decomposition of the field scattered by an array of particles. The particles are reprensented as spheres but the shape can be arbitrary. An electromagnetic plane wave $\textbf{E}_i$ with wavevector $\textbf{k}$ is incident on the metasurface. Current densities $\textbf{J}$ are produced within the particles which radiate a scattered field $\textbf{E}_s$ in both the top and bottom directions. The interference of the incident field with the scattered field produces a reflected ($\textbf{E}_r$) and transmitted field ($\textbf{E}_t$). The current density within a single particle in the array is decomposed on a vector spherical harmonic basis which enables a determination of the multipole coefficients $a_{n,m}$ and $b_{n,m}$ of the unit cell.}
    \label{fig1}
\end{figure}

The problem we are considering is illustrated on Fig. \ref{fig1}, which depicts a two-dimensional infinite periodic array of particles of period $a$. Particles are  represented as spheres, although the formalism is applicable to particles of arbitrary shape. The array is excited by a plane electromagnetic wave $\textbf{E}_i$ with wavevector $\textbf{k}$. Currents are generated in the volume of each particle due to both the interaction of the incoming wave with the particle as well as its interaction with the wave scattered by all neighbouring particles. These currents are source fields in Maxwell's equations. Each particle emits a scattered field $\textbf{E}_s$ in the positive and negative directions. The superposition of all scattered fields in the negative $z$ direction produces a reflected plane wave $\textbf{E}_r$, while the superposition of all scattered fields in the positive $z$ direction interferes with the incoming field and produces a transmitted plane wave $\textbf{E}_t$. The current density field in the unit cell can then be expended using the spherical basis allowing an exact description of the fields with a full account of multipole orders. This enables particles to be both electrically and magnetically polarizable and to sustain multiple resonances (dipole, quadrupole, octupole, etc...). The surrounding medium is considered to be vacuum in what follows for simplicity, but any host medium to the array is applicable.

With these assumptions, the reflection and transmission coefficients of the array can be derived from the expression of the electric field scattered by a single particle of the lattice. Kruk \textit{et al.} have derived the relationship between the field emitted by an array and the field scattered by a single particle or unit cell of the lattice, using Green's function \cite{kruk2016invited}
\begin{equation}\label{eq:Earray Escatt}
\mathbf{E}_{a}(\mathbf{z})=\dfrac{2 \pi i |\mathbf{z}|}{kS} \mathbf{E}_{s}(\mathbf{z}) 
\end{equation}
where $S=a^2$ is the unit cell area of the array, $\mathbf{E}_{a}$ is the field emitted by the array of particles, $\mathbf{E}_{s}$ is that emitted by a unit cell, and $k$ is the free-space wavevector. This expression is valid when considering far-field scattering and for a subwavelength square lattice spacing only. Next we expand the field emitted by a unit cell onto vector spherical harmonics using spherical coordinates. The full derivation is available in the appendix (see subsection \textit{Reflection and transmission coefficients}), but in the interest of conciseness, we directly provide the expressions of the specular reflection and transmission coefficients in terms of symmetric or even ($\mathcal{E}_n$) and anti-symmetric or odd ($\mathcal{O}_n$) multipoles of order $n$.  
\begin{eqnarray}
    \label{eq:ref_general}
    r&=&-\dfrac{\pi}{2k^2 S}\sum^\infty_{n=1} \left[\mathcal{E}_n-\mathcal{O}_n\right]\\
    \label{eq:tr_general}
    t &=& 1-\dfrac{\pi}{2k^2 S}\sum^\infty_{n=1} \left[\mathcal{E}_n+\mathcal{O}_n\right]
 \end{eqnarray}
where the even and odd multipoles can be expressed as a function of the multipole coefficients as follows
\begin{eqnarray}
    \mathcal{E}_n &=&\sum_{m=-1}^{+1} \left[m(4n+1) a_{2n,m} + (4n-1) b_{2n-1,m} \right]\\
    \mathcal{O}_n &=& \sum_{m=-1}^{+1} \left[m(4n-1) a_{2n-1,m} + (4n+1)  b_{2n,m}\right]
\end{eqnarray}
Writing the coefficients in this way is quite convenient as a clear distinction can be made in the role of odd and even modes in the reflection and transmission. This role was already underlined by Kruk \textit{et al.} in the so-called \textit{generalized Kerker} regime that enables highly transmitting arrays \cite{kruk2016invited}. We also note that similar formulas had already been used and discussed in the case of arrays of cylinders by using a cylindrical harmonic decomposition \cite{liu2017generalized}.

\section{Test on a monolayer of high index particles}

\begin{figure}[t!]
   \centering
   \includegraphics[width=0.6\textwidth]{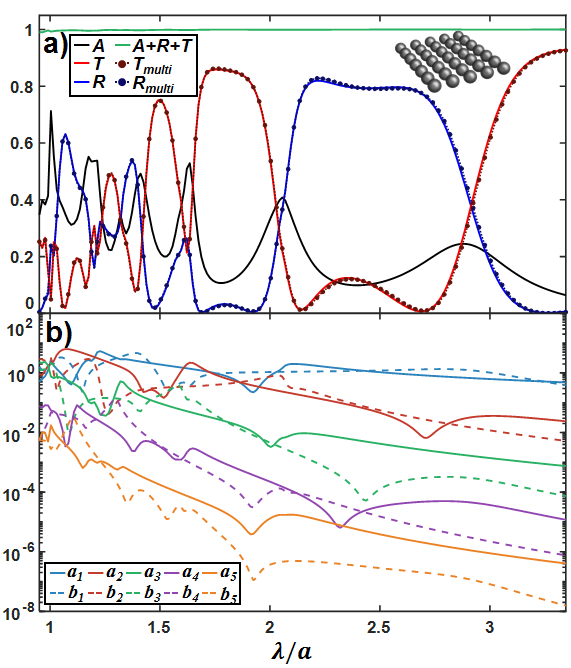}
   \caption{Test of the multipole decomposition. (a) Reflectance (R), transmittance (T) and absorption (A) spectra of an infinite square periodic array of period $a$ composed of a spherical particle of index $N = 4+0.05i$ shown in the inset. The surface fill fraction is 0.4. The continous lines are the spectra calculated from the input and output ports, while the dotted curves are computed from the current decompositions inside the particle and introduced in eqs. \ref{eq:ref_general} and \ref{eq:tr_general}. The absorption is caculated directly from the currents. The green curve equal to 1 across the spectrum is $R+T+A$ and ensures energy conservation. (b) Spectra of the magnitude of the multipolar coefficients. Continuous (dashed) lines represent the electric (magnetic) multipoles.}
    \label{fig2}
\end{figure}
To illustrate this formalism, we apply it to an infinite square array composed of spherical particles made of a fictitious material with loss that has an index of refraction $N=4+0.05i$. The surface fill fraction is taken equal to 0.4. The reflectance ($R$) and transmittance ($T$) spectra of the array are represented on Fig. 2(a) as a function of the normalized wavelength ($\lambda/a$) and are calculated using eqs. \ref{eq:ref_general} and \ref{eq:tr_general} after decomposing the current fields obtained numerically using the eletromagnetics module of the finite-element based software COMSOL Multiphysics. These quantities are compared to the reflectance and transmittance calculated from the input and output ports in COMSOL through a reflected and transmitted planewave projection and are in excellent agreement with one another. Figure 2(b) represents the spectra of the modulus of each multipole coefficients up to the fifth order that are sufficient to correctly reproduce $R$ and $T$. The coefficients shown are the reduced coefficients that are applicable for  symmetry reasons ($a_n=a_{n,1}=-a_{n,-1}$ and $b_n=b_{n,1}=b_{n,-1}$). The absorption ($A$) spectrum is also calculated integrating the ohmic losses in the volume of the particles and the total energy balance is verified by ensuring $R+T+A = 1$. The example of an array composed of particles that exhibit no symmetry is given in the appendix (see subsection \textit{Particle of arbitrary symmetry}).

\section{Application to Perfect Absorption}

Next, we apply the formalism to the problem of complete absorption of a plane wave under normal incidence on a sheet of particles. Many perfect absorbers have been proposed in recent years that are systematically interpreted in terms of critical dipole coupling. The conditions for perfect absorption with an array of dipoles are given by numerous authors \cite{ra2015thin, Alaee_2017}. These conditions can be generalized to a multipolar system using eqs. \ref{eq:ref_general} and \ref{eq:tr_general}. Complete absorption occurs when both the reflection and transmission coefficients are simultaneously cancelled, that is when
\begin{equation}\label{eq:PAcond}
    \sum_{n=1}^{+\infty} \mathcal{O}_n = \sum_{n=1}^{+\infty} \mathcal{E}_n =\frac{k^2S}{\pi}
\end{equation}
This last equation contains two distinct conditions:
\begin{enumerate}
    \item \textbf{The generalized Kerker condition}. As already noticed by \cite{kruk2016invited} in the context of hightly transmitting metasurfaces, this occurs when odd modes are equal to even modes, which we translate here in terms of multipole coefficients as $\sum \mathcal{O}_n =\sum\mathcal{E}_n$. When this condition is realized, we have $r=0$. It should be noted that this general condition only requires the sums of odd and even modes to be equal. It does not require that for every $n$, each symmetric mode be equal to the anti-symmetric odd mode, which would be a much stronger requirement. 
    \item \textbf{The zero transmission condition}. This occurs when $\sum[\mathcal{E}_n+\mathcal{O}_n]=2k^2S/\pi$. Since the relation needs to be satisfied in the complex plane it actually means that the overall sum of odd and even modes needs to be a purely real number and reach the exact value that enables a destructive interference with the incoming plane wave. 
\end{enumerate}

\begin{figure*}[ht!]
   \centering
   \includegraphics[width=\textwidth]{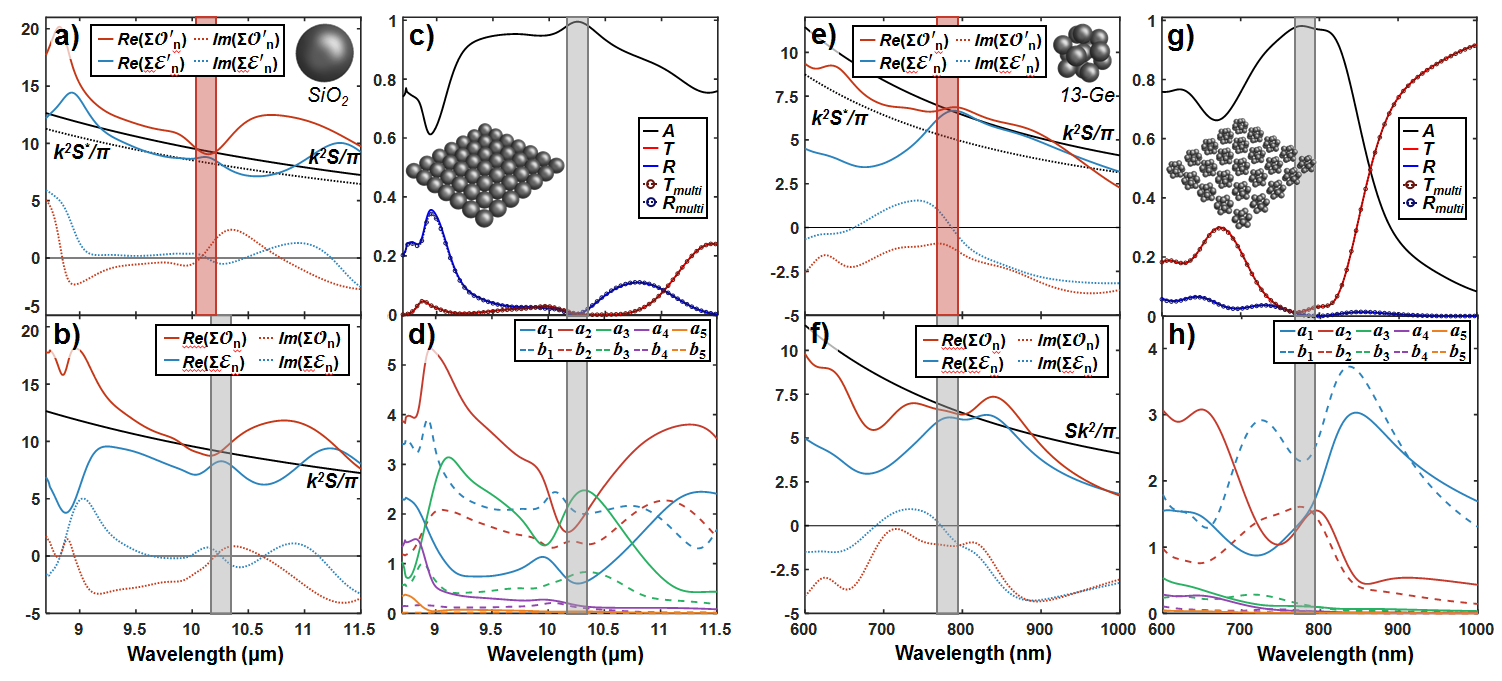}
   \caption{Design of perfect absorbers using the multipolar behavior of individual particles. (a) Wavelength dependence of the sum of even and odd multipole coefficients of an individual SiO$_2$ microsphere in vacuum that is 4.12 $\mu$m in radius, calculated using Mie theory.
   (b) Wavelength dependence of the sum of even and odd multipole coefficients of the SiO$_2$ microsphere in an infinite square array with a surface fill fraction of 0.7. 
   (c) Spectra of $R$, $T$ and $A$ calculated for the square array of SiO$_2$ microspheres, revealing the perfect absorption behavior calculated using the same methods as those used for Fig. 2. (d) Spectra of the modulus of the multipole coefficients retrieved from the current decomposition for the SiO$_2$ microsphere array. 
   (e-h) are respectively the same quantities as those reprensented on panels (a-d) but for a particle cluster composed of thirteen Ge particles that are 75 nm in redius and are pseudo-homgenously distributed in a sphere through repulsive interaction, with a volume fraction of 0.35. The square array made of these clusters has a surface fill fraction of 0.6. The individual cluster was calculated and explored using the T-matrix method.}
    \label{fig3}
\end{figure*}

Equation \ref{eq:PAcond} translates into a \textit{critical coupling} condition equivalent to the classical dipolar critical coupling condition \cite{Alaee_2017}. The power scattered by odd modes is equal to that scattered by even modes and the total scattered power is equal to the absorbed power (see appendix in subsection \textit{Critical coupling equivalent formulation with multipole coefficients and optical theorem}). An equivalent formulation is that
\begin{equation}
    \sigma_s^\mathcal{O}=\sigma_s^\mathcal{E}=\frac{\sigma_a}{2}
\end{equation}
where $\sigma_s^\mathcal{O}$ ($\sigma_s^\mathcal{E}$) and $\sigma_a$ are respectively the scattering cross-sections of odd (even) modes and the absorption cross-section of a particle within the array. We find two of the main features of perfect absorption, that the total scattering is equal to the absorption and that even and odd modes contribute equally and to half of the total scattering each.

Under the assumption that the multipoles of a single particle are only weakly perturbed by the array, this multipolar description becomes very powerful for design as an initial parameter exploration may be done on an individual particle before resorting to a periodic calculation. To illustrate this concept, we shall now propose two designs of perfect absorbers composed of a periodic arrangement of particles that operate in a multipolar regime in contrast to most absorbers investigated thus far \cite{odebo2017large,asadchy2015broadband,liu2017experimental,tian2018near,valagiannopoulos2014symmetric,ming2017degenerate}. First we consider a homogenous microsphere made of SiO$_2$ with a radius $r= 4.12 \mu$m. The sphere is initially considered individually through Mie theory and $r$ is tuned in the mid-infrared region to exploit the phonon resonance that occurs near 10 $\mu$m. The size is chosen to satisfy the \textit{generalized Kerker condition} and make sure that the sum of odd modes and the sum of even modes are real and larger than $k^2S^*/\pi$, where  $S^*=(2r)^2$. This last requirement may be considered a threshold to reach the \textit{zero transmission condition} when the particle will be placed in the array as it would correspond to a hypothetical dense square array. We see from Fig. 3(a) that the conditions are reached since both $\sum \mathcal{O}_n^\prime$ and $\sum  \mathcal{E}_n^\prime$ (priming denotes that the coefficients are those of the isolated particle) become purely real and go beyond the $k^2S^*/\pi$ value near 10.1 $\mu$m in the red shaded area. Next, an array of these particles is simulated and the pitch is fixed to find a value of $S$ that will satisfy eq. \ref{eq:PAcond} as can be seen on Fig. 3(b). This condition is fulfilled for an array with a fill fraction $\pi r^2/S=0.7$ . We note that the array influences both the amplitude of the scattering as well as the wavelength at which the conditions are satisfied in comparison to the isolated particle. 
Figure 3(c) shows the spectra of $R$, $T$, and $A$ of the array. As for Fig. 2(a), the parameters calculated directly from COMSOL through planewave projections are compared to those calculated using eqs. \ref{eq:ref_general} and \ref{eq:tr_general}. Perfect absorption is reached at a wavelength close to 10.2 $\mu$m.  We see from the multipole decomposition in the array (Fig. 3(d)) that the perfect absorption condition given by eq. \ref{eq:PAcond} is almost perfectly preserved showing that our assumption is reasonable and our design strategy is valid. As revealed by Fig. 3(d), we find that the system operates in a heavily multipolar regime (see gray region), where at least 8 multipoles are required to account for the perfect absorption behavior. To our knowledge, no multipolar perfect absorber has been proposed up to now.   
To further reveal the generality of the concept, we also consider a cluster built with spherical particles following an assembly method proposed in \cite{dezert2017isotropic}. Such clusters offer an interesting alternative to spheres made of a homogenous material for which the optical properties are limited by the dispersion of the material. Tuning the volume fraction of inclusions within the clusters enables a tuning of the effective parameters of the cluster. Figure 3(e-h) shows the same quantities as those calculated for the silica system and are computed for a particle cluster composed of thirteen Ge particles that are 75 nm in radius and are pseudo-homogenously distributed in a sphere through repulsive interaction, with a volume fraction of 0.35. The square array made of these clusters has a surface fill fraction of 0.6. Here the resonator is designed to operate at visible frequencies. We see that near perfect absorption is reached at a wavelength close to 790 nm in a heavily multipolar regime once again.

\section{Conclusion}

Through a multipolar description of the reflection and transmission coefficients of a monolayer of arbitrary particles, we show that it is possible to infer properties of the metasurface by directly considering heavily multipolar individual particles. We have applied the formalism introduced to perfect absorption theory and given the generalized condition for total absorption of a plane electromagnetic wave under normal incidence on the surface. This phenomenon occurs when the superposition of odd multipoles is equal to that of even multipoles and are equal to the purely real number $k^2S/\pi$, where $k$ is the free-space wavevector of the impinging wave and $S$ is the surface area of a unit period in the array. We use this result to propose two designs of perfect absorbers, one that operates in the mid-infrared region and is made of a infinite periodic array of resonant silica microspheres and another that operates at visible frequencies and is made of an infinite periodic array of germanium clusters. These results are of great interest to the metasurfaces community that has based a lot of its designs on dipolar and sometimes weakly mutlipolar systems. Furthermore it may interest the self-assembly and nanochemistry community who separate the fabrication of individual particles from the two-dimensional or three-dimensional assembly of materials \cite{rozin2015colloidal,gomez2016hierarchical,baron2016self,ponsinet2017self}. 

\section*{Acknowledgements}
The authors aknowledge financial support from the LabEx AMADEus (ANR-10-LABX-42) in the framework of the IdEx Bordeaux (ANR-10-IDEX-03-02).

\section*{Appendix}

\subsection*{Reflection and transmission coefficients}

In this section we provide the derivation of eqs. \ref{eq:ref_general} and \ref{eq:tr_general}, i.e. the decomposition of $r$ and $t$ as a function of vector spherical harmonic coefficients. 
\medbreak
We consider $\mathbf{E}_{i}$, an $\mathbf{e}_x$ polarized plane wave of magnitude $E_0$ propagating along the $\mathbf{e}_z$ direction, normally incident on the metasurface consisting in a square array of particles. Using a multipolar decomposition and spherical coordinates we want to express $\mathbf{E}_{s}(\mathbf{z})$ the field scattered by a single particle of the array along the $\mathbf{e}_z$ axis for both the forward $(\theta=0,\phi=0)$ and backward $(\theta=\pi,\phi=\pi)$ directions , where $\phi$ is the azimuthal angle and $\theta$ the polar one. For these two directions, due to the polarization of the incident wave, $\mathbf{E}_{s}$ has only one non-zero component along $\mathbf{e}_{\theta}$.
Our starting point is therefore the expression of the $\theta$-component of the scattered field given in \cite{grahn2012electromagnetic} expanded in terms of spherical vector wave functions that takes the following form in the far field domain:
\begin{align}\label{eq:expansion}
    E_{s}(\theta,\phi)=&iE_0\dfrac{e^{ikr}}{kr}\sum_{n=1}^{\infty}\sum_{m=-n}^{+n}\sqrt{\dfrac{\pi(2n+1)}{n(n+1)}}\mathcal{Q}_{nm} \nonumber \\& \times \left[\tau_{nm}(\cos{\theta})a_{nm}+\pi_{nm}(\cos{\theta})b_{nm}  \right]e^{im\phi}
\end{align}
where $\mathcal{Q}_{nm}$ is the normalization constant related to the associated Legendre functions:
\begin{equation}
    \mathcal{Q}_{nm}=\sqrt{\dfrac{2n+1}{4\pi}\dfrac{(n-m)!}{(n+m)!}}
\end{equation}
$\tau_{nm}$ and $\pi_{nm}$ are the angular scattering functions. They can be expressed using the associated Legendre functions $P_n^m(\cos\theta)$ \cite{doicu2006light}:
\begin{align}
    \pi_{nm}(\cos{\theta})&=\dfrac{m}{\sin\theta}P_n^m(\cos\theta) \\
    \tau_{nm}(\cos{\theta})&=\dfrac{d}{d\theta}P_n^m(\cos\theta)
\end{align}
The angular scattering function evaluated for $\theta=0 $ and $\theta=\pi$ can for example be found in \cite{xu1995electromagnetic} and read:
\begin{equation}
\pi_{n,m}(\pm 1)=
\left \{
\begin{aligned}
    &(\pm 1)^{n-1}\dfrac{n(n+1)}{2}  &m=1 \\
    &\dfrac{(\pm 1)^{n-1}}{2}   &m=-1 \\
    &0 &otherwise \\
\end{aligned} 
\right.
\end{equation}

\begin{equation}
\tau_{n,m}(\pm 1)=
\left \{
\begin{aligned}
    &(\pm 1)^{n}\dfrac{n(n+1)}{2} \qquad &m=1 \\
    &-\dfrac{(\pm 1)^{n}}{2}   &m=-1 \\
    & 0 &otherwise \\
\end{aligned} 
\right.
\end{equation}
The angular scattering functions cancel for all $m\notin \{1,-1\}$ for both the forward and backward directions. Using these values, the scattered  electric field reads in the forward direction:
\begin{equation}\label{Escattfor}
    E_{s}(0,0)=i \dfrac{E_0}{4}\dfrac{e^{ikr}}{kr}\sum_{n=1}^{\infty}(2n+1) \left[ a_{n,1}-a_{n,-1}+b_{n,1} +b_{n,-1} \right]
\end{equation}
and in the backward direction:
\begin{equation}\label{Escattback}
    E_{s}(\pi,\pi)=i \dfrac{E_0}{4}\dfrac{e^{ikr}}{kr}\sum_{n=1}^{\infty}(2n+1)(-1)^{n+1} \left[ a_{n,1} -a_{n,-1} -b_{n,1} -b_{n,-1}  \right]
\end{equation}
Combining \ref{eq:ref_general} and \ref{eq:tr_general} with \ref{eq:expansion}, the field leaving the array can be expressed:
\begin{equation}\label{eq:Etrans}
    E_{a}(z_+) = -\dfrac{\pi}{2k^2 S}\sum^\infty_{n=1} (2n+1) \left( a_{n,1}-a_{n,-1} \right. \left. +b_{n,1} +b_{n,-1} \right){E}_{0}
\end{equation}
\begin{equation}
    E_{a}(z_-) =-\dfrac{\pi}{2k^2 S}\sum^\infty_{n=1} (-1)^{n+1} (2n+1) \left( a_{n,1}-a_{n,-1} -b_{n,1} -b_{n,-1} \right) {E}_{0}
\end{equation}
Inserting these 2 expressions in $\mathbf{E}_t=\mathbf{E}_{i}+\mathbf{E}_{a}(z_+)$ and $\mathbf{E}_r=\mathbf{E}_{a}(z_-)$ allows us to finally write compactly the reflection and transmission coefficients expressed as a function of the spherical multipole coefficients of a single resonator (or unit cell) of the array:
\begin{align}\label{eq:transmission_general}
    t&=1-\dfrac{\pi}{2k^2 S}\sum^\infty_{n=1} \sum_{m=-1,+1} (2n+1) \left( ma_{n,m}+b_{n,m}\right) \\
    r&=-\dfrac{\pi}{2k^2 S}\sum^\infty_{n=1} \sum_{m=-1,+1} (-1)^{n+1} (2n+1) \left( ma_{n,m}-b_{n,m}\right)\label{eq:reflexion_general}
\end{align}

\subsection*{Particle of arbitrary symmetry}

\begin{figure}[b!]
   \centering
   \includegraphics[width=0.7\textwidth]{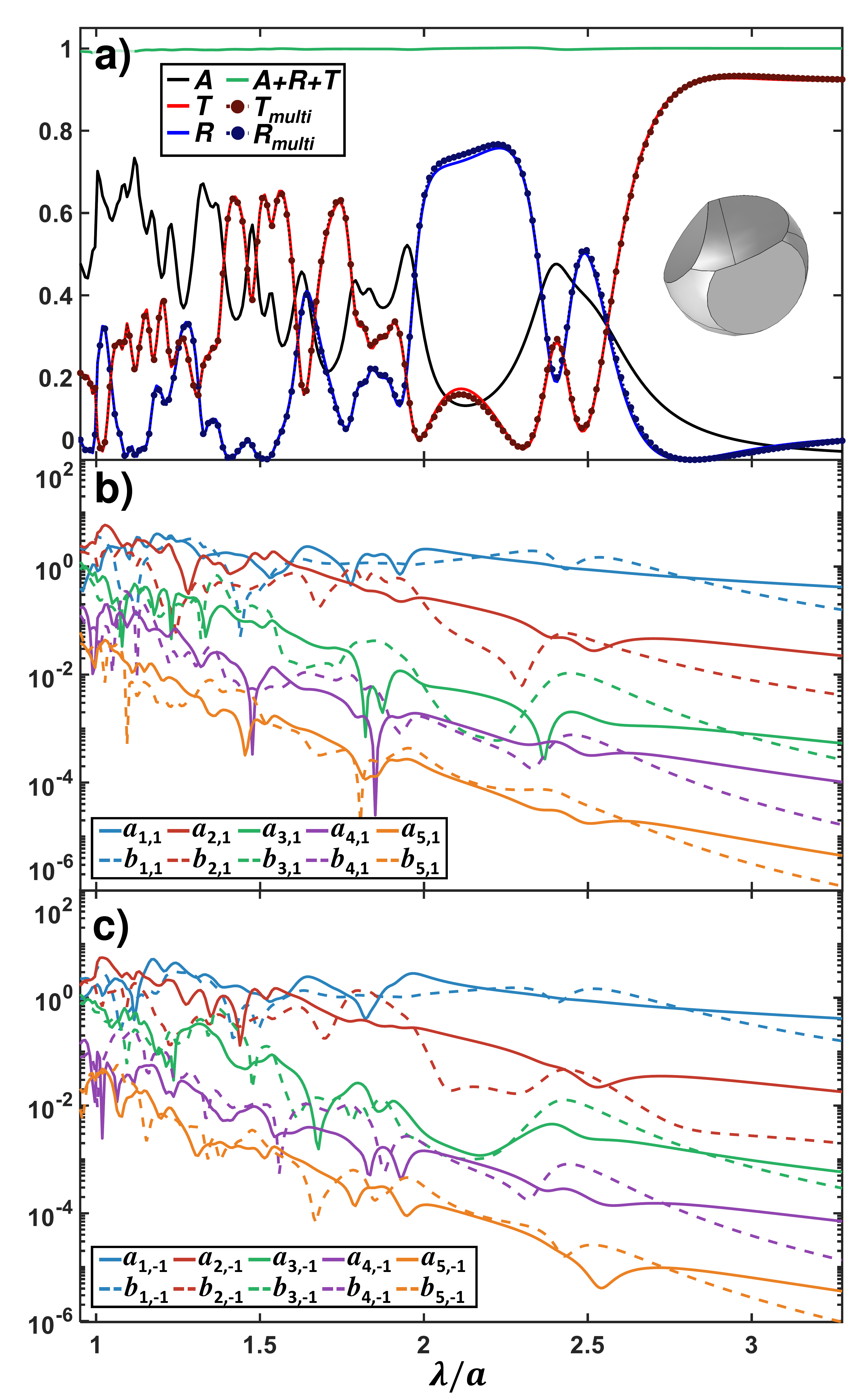}
   \caption{Validation of the multipole decomposition with unsymmetrical particles. (a) Reflectance ($R$), transmittance ($T$) and absorption ($A$) spectra of an infinite square periodic array of period $a$ composed of a 3 times truncated spherical particle. The continuous lines are the spectra of the specular reflection and transmission calculated from the input and output ports, while the dashed curves are computed from the extracted multipoles as introduced in eqs. \ref{eq:reflexion_general} and \ref{eq:transmission_general}. The green curve equal to 1 across the spectrum is $R+T+A$ ensuring energy conservation. (b) Spectra of the magnitude of the multipole coefficients of order $m=1$. Continuous (dashed) lines represents the electric (magnetic) multipoles.(c) Spectra of the magnitude of the multipole coefficients of order $m=-1$.}
    \label{figsup1}
\end{figure}

The introduced multipole-based formalism can be used to study periodic arrays of particles with arbitrary shape. To illustrate its large domain of validity, we apply it to arrays of particles consisting of truncated spherical particles with refractive index $N=4+0.05i$ and radius $r=150$ nm. Truncations by 2 random planes and a curved surface were used to completely break the symmetry of the final structure as illustrated in the inset of Fig. \ref{figsup1}(a). The surface fill fraction is taken equal to 0.4 as in Fig. 2 in the main text. The specular reflectance ($R$) and transmittance ($T$) as well as the absorption spectra ($A$) of the array are represented on Fig. \ref{figsup1}(a) as a function of the normalized wavelength ($\lambda/a$). These quantities are well reproduced by the reflectance and transmittance calculated using the extracted multipoles with eqs. 2 and 3 in the main text.
The spectra of the modulus of each multipole coefficients of order $m=1$ ($m=-1$) up to the fifth degree are represented on Figure \ref{figsup1}(b) (Figure \ref{figsup1}(c)).

\subsection*{Critical coupling equivalent formulation with multipole coefficients and optical theorem}

In this section we provide the derivation of eq.7 i.e. the energy balance for the metasurface translating the critical coupling of both the symmetric and anti-symmetric excited modes. 
\medbreak
By definition, the fraction of incoming power scattered in the backward (forward) direction by the array is equal to $\lvert r\rvert^2$ ($\lvert t-1\rvert^2$). 
Using these definitions and the expressions of the reflection and transmission coefficients (i.e. eqs. \ref{eq:transmission_general} and \ref{eq:reflexion_general}) the scattering cross sections of even and odd multipoles of a unit cell within the array may be defined by multiplying those quantities by the surface area of a unit cell $S$, which yields
\begin{align}
    \sigma_s^\mathcal{E}&=\dfrac{\pi^2}{2k^4S}\left|\sum^\infty_{n=1} \mathcal{E}_n\right|^2\\
\sigma_s^\mathcal{O}&=\dfrac{\pi^2}{2k^4S}\left|\sum^\infty_{n=1} \mathcal{O}_n\right|^2
\end{align}
Assuming perfect absorption, with the condition $\sum_{n=1}^{+\infty} \mathcal{O}_n = \sum_{n=1}^{+\infty} \mathcal{E}_n =\frac{k^2S}{\pi}$, the two contributions simply become
\begin{equation}\label{eq:crical2}
    \sigma_s^\mathcal{O}=\sigma_s^\mathcal{E}=\frac{\sigma_{a}}{2}
\end{equation}
where the absorption cross-section is actually equal to $S$ since all of the energy impinging a unit cell is absorbed. 

\medbreak
We can also show that the optical theorem holds by considering the total absorption $A=1-R-T$, which can be expressed with the multipole coefficients and provides the proportion of absorption due respectively to the symmetric and anti-symmetric modes. Using eqs. \ref{eq:reflexion_general} and \ref{eq:transmission_general} the energy balance then reads:
\begin{equation}
    A=-2{\left(\dfrac{\pi}{2k^2S}\right)}^2 \left[\left|\sum^\infty_{n=1} \mathcal{E}_n\right|^2+\left|\sum^\infty_{n=1} \mathcal{O}_n\right|^2 \right]+ \dfrac{\pi}{k^2S}\sum^\infty_{n=1} Re\left(\mathcal{E}_n+\mathcal{O}_n\right)
\end{equation}
or equivalently
\begin{equation}\label{eq:bilan}
    \sigma_{a}=-\sigma_s+\sigma_{ext}
\end{equation}
where $\sigma_{ext}$ is the extinction cross section of the particles in the array, which in the case of an $x$-polarized incident plane wave becomes \cite{grahn2012electromagnetic}:
\begin{equation}\label{eq:extcalcul}
\sigma_{ext}=\dfrac{\pi}{k^2}\sum_{n=1}^\infty\sum_{m=-1,+1}(2n+1)Re\left( Ma_{n,m}+b_{n,m} \right)
\end{equation}

\begin{figure}[t!]
   \centering
   \includegraphics[width=0.7\textwidth]{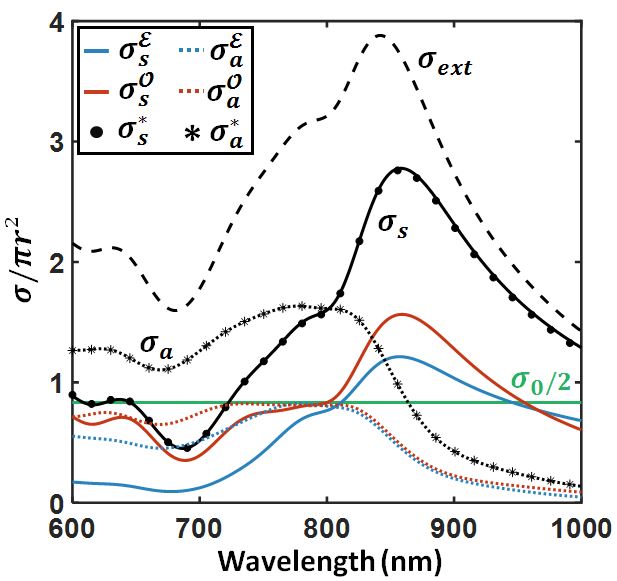}
   \caption{Cross section decomposition in terms of even and odd contributions apply to a cluster-based absorbing metasurface. Each cluster consist of thirteen Ge particles that are 75 nm in radius distributed in a sphere of radius $R=250$nm. The square array has a surface fill fraction of 0.6. 
}
    \label{figsup2}
\end{figure}

As for the scattering cross section, the extinction cross section expression can be broken down into even and odd contributions as:
\begin{align}
\sigma_{ext}^\mathcal{E}&=\dfrac{\pi}{k^2}\sum^\infty_{n=1} Re(\mathcal{E}_n) \\
\sigma_{ext}^\mathcal{O}&=\dfrac{\pi}{k^2} \sum^\infty_{n=1} Re(\mathcal{O}_n)
\end{align}
In the same way the absorption cross section can be decomposed into even and odd terms obtained by the differences
\begin{align}
\sigma_{a}^\mathcal{O}&=\sigma_{ext}^\mathcal{O} - \sigma_s^\mathcal{O}\\
\sigma_{a}^\mathcal{E}&=\sigma_{ext}^\mathcal{E} - \sigma_s^\mathcal{E}
\end{align}
We see that the optical theorem holds for odd and even superpositions independently.

Accessing this decomposition would usually require a configuration with double illumination from both sides of the array as described in \cite{ming2017degenerate} with co- or cross-polarization allowing for a selective excitation of either the symmetric or anti-symmetric modes of the structure. Instead, using the multipole decomposition allows for a one step procedure to retrieve both the emission and absorption due to the even or odd excited modes.

To illustrate this approach, we present in Fig. \ref{figsup2} the decomposition of the scattering and absorption cross-sections in terms of even and odd multipoles applied to the Ge cluster-based absorber presented in the main text. The absorption $\sigma_a^*$ and emission $\sigma_s^*$ cross section calculated with respectively the absorbed power obtained via the integration of the ohmic losses and the reflection and transmission coefficients are perfectly reproduced by those reconstructed from the multipole coefficients. In the range of high absorption, the cross sections satisfy:
\begin{equation}
\left\{
\begin{aligned}
        \sigma_s^{\mathcal{O}} &\approx \sigma_a^{\mathcal{O}}\approx \dfrac{\sigma_0}{2} \\
    \sigma_s^{\mathcal{E}} &\approx \sigma_a^{\mathcal{E}}\approx \dfrac{\sigma_0}{2}
\end{aligned}
 \right.
\end{equation}
where $\sigma_0=S$. Strict equalities would be obtained in the case of an absorption reaching $100\%$ while in the present case the maximum absorption exhibited by the metasurface is $98\%$. 

The presented cross sections are quantities of interest as they carry the same information than the scattering, extinction and absorption cross section defined for isolated particles in homogeneous space \cite{jackson1999classical,grahn2012electromagnetic}. Therefore they allow for a direct comparison between the optical properties of particles isolated or placed in a periodic array.

\bibliographystyle{unsrt}  


\end{document}